\documentclass[peerreview]{IEEEtran}

\usepackage[margin=1in]{geometry}
\usepackage[utf8]{inputenc}
\usepackage{amsmath,bm}
\usepackage{amsthm}
\usepackage{amssymb}
\usepackage{mathtools}
\usepackage{bbm}
\allowdisplaybreaks
\usepackage{graphicx}
\usepackage[dvipsnames]{xcolor}
\usepackage[english]{babel}
\usepackage[font=small]{caption}
\usepackage{subcaption}
\usepackage{wrapfig}
\usepackage{tabularx}
\usepackage{multirow}
\usepackage{multicol}
\usepackage{thmtools, thm-restate}
\usepackage[normalem]{ulem}
\usepackage[noadjust]{cite}
\usepackage{makecell}
\usepackage{cancel}
\usepackage{hyphenat}
\usepackage{enumitem}
\usepackage[linesnumbered,ruled,vlined]{algorithm2e}
\SetKwInput{KwInput}{Input}                
\SetKwInput{KwOutput}{Output}              
\usepackage{algorithmic}

\usepackage[final]{changes}

\newtheorem{theorem}{Theorem$\!$}
\newtheorem{claim}{Claim$\!$}
\newtheorem{example}{Example$\!$}
\newtheorem{lemma}{Lemma$\!$}
\newtheorem{corollary}{Corollary$\!$}
\newtheorem{proposition}{Proposition$\!$}

\newtheorem{definition}{Definition$\!$}
\usepackage{graphicx}

\usepackage{xcolor}

\newcommand{\cB}{\mathcal{B}}
\newcommand{\cC}{\mathcal{C}}
\newcommand{\cD}{\mathcal{D}}

\newcommand{\cG}{\mathcal{G}}

\newcommand{\cM}{\mathcal{M}}

\newcommand{\cP}{\mathcal{P}}

\newcommand{\cS}{\mathcal{S}}

\newcommand{\mybold}[1]{\bm{#1}}

\newcommand{\bsf}{{\mybold{f}}}
\newcommand{\bg}{{\mybold{g}}}

\newcommand{\bp}{{\mybold{p}}}

\newcommand{\br}{{\mybold{r}}}

\newcommand{\bx}{{\mybold{x}}}
\newcommand{\by}{{\mybold{y}}}

\newcommand{\balpha}	{\mybold{\alpha}}

\newcommand{\bpi}		{\mybold{\pi}}

\newcommand{\bsigma}	{\mybold{\sigma}}
\newcommand{\btau}		{\mybold{\tau}}

\IEEEoverridecommandlockouts

\begin{document}


\title{Permutation and Multi-permutation Codes Correcting Multiple Deletions}

\author{\textbf{Shuche Wang}\IEEEauthorrefmark{1},  
\textbf{The Nguyen}\IEEEauthorrefmark{2},
\textbf{Yeow Meng Chee}\IEEEauthorrefmark{3},
and \textbf{Van Khu Vu}\IEEEauthorrefmark{3}\\
\IEEEauthorblockA{
\IEEEauthorrefmark{1}  Institute of Operations Research and Analytics, National University of Singapore, Singapore \\[0.5mm]
\IEEEauthorrefmark{2} Department of Mathematics, University of Illinois Urbana-Champaign, United States \\[0.5mm]
\IEEEauthorrefmark{3}  Department of Industrial Systems Engineering
and Management, National University of Singapore, Singapore\\[0.5mm]
}
{Emails:\, shuche.wang@u.nus.edu, thevn2@illinois.edu, \{ymchee,isevvk\}@nus.edu.sg}

\thanks{ This paper was presented in part at the 2024 IEEE International Symposium on Information Theory (ISIT) in 2024~\cite{wang2024permutation}.}
}

\maketitle


\begin{abstract}
Permutation codes in the Ulam metric, which can correct multiple deletions, have been investigated extensively recently. In this work, we are interested in the maximum size of permutation codes in the Ulam metric and aim to design permutation codes that can correct multiple deletions with efficient decoding algorithms. We first present an improvement on the Gilbert--Varshamov bound of the maximum size of these permutation codes by analyzing the independence number of the auxiliary graph. The idea is widely used in various cases and our contribution in this section is enumerating the number of triangles in the auxiliary graph and showing that it is small enough. 
Next, we design permutation codes correcting multiple deletions with a decoding algorithm. In particular, the constructed permutation codes can correct $t$ deletions with at most $(3t-1) \log n+o(\log n)$ bits of redundancy where $n$ is the length of the code. Our construction is based on a new mapping  which yields a new connection between permutation codes in the Hamming metric and permutation codes in various metrics. 
Furthermore, we construct permutation codes that correct multiple bursts of deletions using this new mapping.
Finally, we extend the new mapping for multi-permutations and construct the best-known multi-permutation codes in Ulam metric. 
\end{abstract}
 
\section{Introduction}
Permutation codes in the Hamming metric were introduced by Slepian~\cite{slepian1965permutation} in 1965 for transmitting data in the presence of additive Gaussian noises. Since then, they have been extensively studied for their effectiveness in powerline transmission systems against impulsive noise~\cite{chu2004constructions}, as well as in the development of block ciphers~\cite{de2000application}. Besides that, permutation codes in various metrics, including Kendall-$\tau$ metric~\cite{buzaglo2014perfect}, generalised-Kendall-$\tau$ metric\cite{chee2014breakpoint}, 
generalised-Cayley metric~\cite{chee2014breakpoint}, Chebyshev metric~\cite{klove2010permutation}, Ulam metric~\cite{farnoud2013error}, have attracted a lot of attention from theoretical points of view as well as their application in flash memory with rank modulation~\cite{jiang2009rank,jiang2010correcting,farnoud2013error,barg2010codes}.

Flash memories have emerged as a promising nonvolatile data storage solution, favored for their speed, low power usage, and reliability. To circumvent the precise programming of each cell to a specific level in flash memory devices, Jiang et al.~\cite{jiang2009rank} introduced a rank modulation method that utilizes permutations to represent information. To address the issue of deletions or erasures in this context, Gabrys et al.~\cite{gabrys2015codes} categorized these deletions into two types: symbol-invariant deletions (SID) and permutation-invariant deletions (PID). 

Our goal in this work is to study permutation codes that can correct $t$ deletions under the SID model, where deleting some symbols does not affect the values of others. For simplicity of expression, all \emph{deletions} we refer to in the remainder of this paper represent \emph{symbol invariant deletions}. Although there has been a significant breakthrough in the codes for correcting $t$ deletions in general binary~\cite{sima2020optimalbinary,sima2020syndrome} and non-binary~\cite{sima2020optimal} scheme, permutation codes for correcting $t$ deletions are not as well-studied. Gabrys et al.~\cite{gabrys2015codes} proposed permutation codes for correcting $t$ deletions under the SID model by demonstrating the equivalence with permutation-invariant erasures (PIE) codes, which is facilitated through the integration of a permutation code in the Ulam metric. Farnoud et al.~\cite{farnoud2013error} provided construction for permutation codes in the Ulam metric. Recently, Goldenberg et al.~\cite{goldenberg2024explicit} proposed explicit permutation codes capable of approaching relative distance $1$ in the Ulam metric. Our work focuses on considering the Ulam/Levenshtein distance is constant. Besides, there is another line of work on studying permutation codes for correcting a burst of $t$ deletions~\cite{sun2023improved,Wang2022permutation,chee2019burst}. Order-optimal permutation codes capable of correcting a burst of at most $t$ deletions are proposed in~\cite{sun2023improved,Wang2022permutation}.

Afterward, \emph{multipermutation codes}, which generalize permutation codes have also been incorporated into the rank modulation scheme used in flash memory systems~\cite{gad2012trade,hassanzadeh2014multipermutation,sala2014deletions}. These codes enable more efficient encoding strategies by leveraging the structure of multisets, making them well-suited for memory storage applications. In the context of the Hamming metric, multipermutation codes are referred to as constant composition codes or frequency permutation arrays (FPAs). Several studies~\cite{luo2003constant,ding2005combinatorial,huczynska2006frequency} have explored the properties and applications of these codes. Farnoud and Milenkovic~\cite{hassanzadeh2014multipermutation} presented bounds on the size of multipermutation codes for both the Ulam and Hamming metrics. They also presented constructions for multipermutation codes in the Ulam metric. Sala et al. studied the multipermutation codes for correcting single and multiple deletions in \cite{sala2014deletions}. Also, multipermutation codes for correcting a burst of deletions are also studied in \cite{sun2023improved,han2023constructions}.

In 1965, Levenshtein~\cite{levenshtein1966binary} established the upper and lower bounds of the maximum size of the binary code for correcting $t$ deletions. 
Jiang and Vardy~\cite{jiang2004asymptotic} proposed a technique to obtain an asymptotic improvement on the Gilbert--Varshamov (GV) bounds of binary codes in the Hamming metric. Later, Vu and Wu \cite{vu2005improving} applied the technique to improve the GV bound of $q$-ary codes in the Hamming metric. 
Recently, Alon et al.~\cite{alon2023logarithmically} used the same technique to achieve a logarithmic improvement on the GV bound of the maximum size of the binary $t$-deletion codes. 
For the permutation code, several works studied the improvement on the GV bound of the maximum size of the permutation code of length $n$ in the Hamming metric $d$~\cite{gao2013improvement,tait2013asymptotic,jin2015construction,wang2017new}, in the Cayley metric~\cite{nguyen2024improving}, and Kendall-$\tau$ metric \cite{nguyen2024improving}. In this work, we aim to use the technique proposed by Jiang and Vardy in \cite{jiang2004asymptotic} to improve the GV lower bound of the maximal size of permutation codes correcting $t$ deletions. Specifically, we achieve this by creating an auxiliary graph on the set of permutations. In this graph, each $t$-deletion permutation code corresponds to an independent set. Consequently, determining the size of these codes involves examining the independence number of the auxiliary graph.

 Then, we focus on designing permutation codes in the Ulam metric which can correct multiple deletions. We establish relationships between various kinds of distances over two permutations, including Hamming, Levenshtein, Ulam, and (generalized) Kendall-$\tau$ distance. In particular, the main idea of our work is that we introduce a novel mapping function such that we successfully build a tighter inequality relationship between the Ulam distance and Hamming distance compared with the result in \cite{farnoud2013error}. It helps us to construct permutation codes for deletions/translocations/transpositions by applying the well-studied permutation codes in the Hamming metric for substitutions as the base code. In particular, we show that there exist permutation codes of length $n$ for correcting $t$ deletions with at most $(3t-1) \log n+o(\log n)$ bits of redundancy, where $t$ is a constant. Furthermore, we present the construction of permutation codes for correcting up to $t$  deletions with a specific decoding process. Instead of relying on the auxiliary codes in the Ulam metric~\cite{gabrys2015codes}, our construction is achieved by incorporating the base code in the Hamming metric. The redundancy of our proposed code improves that of in \cite{gabrys2015codes}. Next, our novel mapping function can be considered a useful tool for constructing permutation codes for different types of channels. As an example, we demonstrate its application in designing permutation codes capable of correcting multiple bursts of deletions. Also, we extend the permutation code for correcting $t$ deletions to multipermutation, which is achieved by leveraging our novel mapping function in a more general way.

The remainder of this paper is structured as follows. Section~\ref{sec:notation} presents the notations and preliminaries. Section~\ref{sec:improved_gv} provides an improved GV bound of the $t$-deletion permutation code. In Section~\ref{sec:relation_distance}, we build a tight inequality relationship between the Ulam distance and Hamming distance via a novel mapping function, which helps us to construct permutation codes for correcting $t$ deletions with a lower redundancy. In Section~\ref{sec:construction}, we construct permutation codes for correcting $t$ deletions with a specific decoding process. In Section~\ref{sec:multiburst}, we construct permutation codes for correcting multiple bursts of deletions. In Section~\ref{sec:multipermutation}, we extend the permutation code for correcting $t$ deletions to multipermutation. Finally, Section~\ref{sec:conclusion} concludes the paper.


\section{Notation and Preliminaries}\label{sec:notation}

Given an integer $n$, let $[n]$ denote the set $\{1,2,\ldots,n\}$. A permutation is a bijection $\bsigma: [n] \mapsto [n]$ and denoted $\bsigma =(\sigma(1),\sigma(2),\ldots,\sigma(n))$.
Let $\cS_n$ be the set of all permutations on $[n]$, that is the symmetric group of order $n!$. Denote $\bpi=(\pi_1,\pi_2,\dotsc,\pi_{n})\in \cS_n$ as a permutation with length $n$.  Given a permutation \( \bpi = (\pi_1, \pi_2, \ldots, \pi_n) \in \cS_n \), the inverse permutation is denoted as \( \bpi^{-1} = (\pi^{-1}_1, \pi^{-1}_2, \ldots, \pi^{-1}_n) \). Here, \( \pi^{-1}_i \) indicates the position of the element \( i \) in the permutation \( \bpi \). For an integer $x \in [n]$, $\pi^{-1}(x)$ indicates the position of $x$ in permutation $\bpi$. Also, $\bpi_{[i,j]}$ denotes the subsequence beginning at index $i$ and ending at index $j$, inclusive. For functions, if the output is a sequence, we also write them with bold letters, such as $\bp(\bpi)$. The $i$th position in $\bp(\bpi)$ is denoted $p(\bpi)_i$.
\begin{example}
    Suppose $\bpi=(1,3,4,2,5)$, we have $\bpi^{-1}=(1,4,2,3,5)$ and $\pi^{-1}(2)=4$.
\end{example}

\subsection{Basic Definitions}
\begin{definition}
    For distinct $i,j\in[n]$, a transposition $\tau(i,j)$ leads to a new permutation obtained by swapping $\pi_i$ and $\pi_j$ in $\bpi$, i.e,
    \begin{equation*}
        \bpi\tau(i,j)=(\pi_1,\dotsc,\pi_{i-1},\pi_{j},\pi_{i+1},\dotsc,\pi_{j-1},\pi_{i},\pi_{j+1},\dotsc,\pi_n).
    \end{equation*}
    If $|i-j|=1$, $\tau(i,j)$ is called the \emph{adjacent transposition}.
\end{definition}

\begin{definition}
    For distinct $i,j\in[n]$, if $i< j$, a translocation $\phi(i,j)$ leads to a new permutation obtained by moving $\pi_i$ to the position of $j$ and shifting symbols $\pi_{[i+1,j]}$ by one in $\bpi$. There is 
    \begin{equation*}
        \bpi\phi(i,j)=(\pi_1,\dotsc,\pi_{i-1},\pi_{i+1},\dotsc,\pi_{j},\pi_{i},\pi_{j+1},\dotsc,\pi_n).
    \end{equation*}
    If $i> j$, a translocation $\phi(i,j)$ leads to a new permutation obtained by moving $\pi_i$ to the position of $j$ and shifting symbols $\pi_{[j,i-1]}$ by one in $\bpi$. There is 
    \begin{equation*}
        \bpi\phi(i,j)=(\pi_1,\dotsc,\pi_{j-1},\pi_{i},\pi_{j},\dotsc,,\pi_{i-1},\pi_{i+1},\dotsc,\pi_n).
    \end{equation*}
\end{definition}

In \cite{chee2014breakpoint}, Chee and Vu introduced the \emph{generalized transposition} as follows. Denote $[a,b]\prec[c,d]$ as the interval $[a,b]$ precedes the interval $[c,d]$.
\begin{definition}
    For distinct two intervals $A=[i,j],B=[k,\ell]$, a generalized transposition $\tau_g(A,B)$ leads to a new permutation obtained by swapping $\bpi_{[i,j]}$ and $\bpi_{[k,\ell]}$ in $\bpi$. If $A\prec B$, there is
    \begin{equation*}
        \bpi\tau_g(A,B)=(\pi_1,\dotsc,\pi_{i-1},\pi_{k},\pi_{k+1},\dotsc,\pi_{\ell},\pi_{j+1},\dotsc,\\
        \pi_{k-1},\pi_{i},\dotsc,\pi_{j},\pi_{\ell+1},\dotsc,\pi_n).
    \end{equation*}
    If $k-j=1$, we denote $\tau_a(A,B)$ as the generalized adjacent transposition.
\end{definition}

We have $\phi(i,\ell)=\tau_a(A,B)$ with $A=[i,j],B=[k,\ell]$ if $i=j=k-1$. Hence, we say that a \emph{translocation} can be considered as a special case of a \emph{generalized adjacent transposition}. We note that computing the exact generalized adjacent transposition distance between two permutations is an NP-hard problem.

\begin{example}
    Let $\bpi=(1,6,4,3,2,5)$. We have the following:
    \begin{align*}
        \bpi\tau(2,5)=(1,2,4,3,6,5),\\
        \bpi\phi(2,5)=(1,4,3,2,6,5),\\
        \bpi\tau_g([2,3],[5,6])=(1,2,5,3,6,4),\\
        \bpi\tau_a([2,3],[4,6])=(1,3,2,5,6,4).
    \end{align*}
\end{example}

\begin{definition}
    The Hamming distance between two permutations \( \bpi, \bsigma \in \cS_n \), denoted by \( d_H(\bpi,\bsigma) \), is defined as the number of positions for which \( \bpi \) and \( \bsigma \) differ, i.e,
    \begin{equation*}
        d_{H}(\bpi,\bsigma)=|\{i\in[n]:\pi_i\neq \sigma_i\}|.
    \end{equation*}
\end{definition}

\begin{definition}
    The Levenshtein distance between two permutations \( \bpi, \bsigma \in \cS_n \), denoted by \( d_L(\bpi,\bsigma) \), is defined as the minimum number of insertions or deletions which are needed to change $\bpi$ to $\bsigma$. 
\end{definition}

\begin{definition}
    The Kendall-$\tau$ distance between two permutations \( \bpi, \bsigma \in \cS_n \), denoted by \( d_K(\bpi,\bsigma) \), is defined as the minimum number of adjacent transpositions which are needed to change $\bpi$ to $\bsigma$, i.e,
    \begin{equation*}
        d_{K}(\bpi,\bsigma)=\min\{m:\bpi\tau_1\tau_2\dotsm\tau_m=\bsigma\},
    \end{equation*}
    where $\tau_1,\dotsc,\tau_m$ are adjacent transpositions.
\end{definition}

\begin{definition}
    The generalized Kendall-$\tau$ distance between two permutations \( \bpi, \bsigma \in \cS_n \), denoted by \( d_{\bar{K}}(\bpi,\bsigma) \), is defined as the minimum number of generalized adjacent transpositions which are needed to change $\bpi$ to $\bsigma$, i.e,
    \begin{equation*}
        d_{\bar{K}}(\bpi,\bsigma)=\min\{m:\bpi\tau_{a1}\tau_{a2}\dotsm\tau_{am}=\bsigma\},
    \end{equation*}
    where $\tau_{a1},\dotsc,\tau_{am}$ are generalized adjacent transpositions.
\end{definition}

\begin{definition}
    The generalized Cayley distance between two permutations \( \bpi, \bsigma \in \cS_n \), denoted by \( d_{C}(\bpi,\bsigma) \), is defined as the minimum number of generalized transpositions which are needed to change $\bpi$ to $\bsigma$, i.e,
    \begin{equation*}
        d_{C}(\bpi,\bsigma)=\min\{m:\bpi\tau_{g1}\tau_{g2}\dotsm\tau_{gm}=\bsigma\},
    \end{equation*}
    where $\tau_{g1},\dotsc,\tau_{gm}$ are generalized transpositions.
\end{definition}

\begin{definition}
    The Ulam distance between two permutations \( \bpi, \bsigma \in \cS_n \), denoted by \( d_{U}(\bpi,\bsigma) \), is defined as the minimum number of translocations which are needed to change $\bpi$ to $\bsigma$, i.e,
    \begin{equation*}
        d_{U}(\bpi,\bsigma)=\min\{m:\bpi\phi_{1}\phi_{2}\dotsm\phi_{m}=\bsigma\},
    \end{equation*}
    where $\phi_{1},\dotsc,\phi_{m}$ are translocations.
\end{definition}

\begin{proposition}\label{prop:ulam}
    For two permutations \( \bpi,\bsigma \in \cS_n \), let \( \mathrm{LCS}(\bpi,\bsigma) \) be the length of a longest common subsequence of \( \bpi \) and \( \bsigma \). The Ulam distance \( d_U(\bpi,\bsigma)\) between \( \bpi \) and \( \bsigma \) equals  \( n - \mathrm{LCS}(\bpi,\bsigma) \).
\end{proposition}
\begin{example}
    Let \( \bpi = (4,3,1,2,5) \) and \( \bsigma = (4,3,5,1,2) \). Then \( d_H(\bpi,\bsigma) = 3 \) and \( d_U(\bpi,\bsigma) = 1 \).
\end{example}

\subsection{Permutation Code}
Given a permutation $\bpi\in \cS_n$, let $\cD_{t}(\bpi)$ be the set of all vectors of length $n-t$ received as a result of $t$ deletions in $\bpi$ and let $\cB_{t}(\bpi)\subseteq \cS_n$ denote the \emph{confusable set} of $\bpi$, i.e., the set of permutations $\bsigma$ other than $\bpi$ for which $\cD_{t}(\bpi) \cap \cD_{t}(\bsigma)\neq \emptyset$. Constructing the permuation codes for correcting $t$ deletions is to design codes $\cP_{t}(n)$ over $\cS_n$ where for any $\bpi \in \cP_{t}(n)$, we can recover $\bpi$ from $\bpi'$, provided that $\bpi'$ is the result of $t$ deletions occurring in $\bpi$. The code $\cP_{t}(n)$ is a $t$-deletion permutation code if and only if for distinct $\bpi_1,\bpi_2\in \cP_{t}(n)$ such that $\cD_{t}(\bpi_1)\cap \cD_{t}(\bpi_2)=\emptyset$. The size of a permutation code $\cC\subseteq \cS_n$ is denoted $|\cC|$ and its redundancy is defined as $\log (n!/|\cC|)$. All logarithms in this paper are to the base 2.

In \cite{levenshtein1992perfect}, Levenshtein demonstrated that for sequences $\bx$ and $\by$ with a length of $n$, the Levenshtein distance between $\bx$ and $\by$ satisfies $d_{L}(\bx,\by)=2(n-\mathrm{LCS}(\bx,\by))$. This formula applies to permutations as well. Combining with the Proposition~\ref{prop:ulam}, we have:
\begin{equation}\label{eq:ulamleven}
    d_{L}(\bpi,\bsigma)=2 d_U(\bpi,\bsigma).
\end{equation}
Therefore, a \textit{$t$-deletion permutation code} of length $n$ is a subset $\cC$ of $\cS_n$ such that the Ulam distance of any pair of distinct elements in $\cC$ is at least $t+1$. We would like to determine the maximum size $M(n, t)$ of a $t$-deletion permutation code. 

\begin{lemma}
Let $n>t$ be positive integers. $M(n, t)$ is bounded as follows
\begin{equation}
    \Omega_t\left(\frac{n!}{n^{2t}}\right) \leq M(n,t) \leq O_t\left(\frac{n!}{n^t}\right).
\end{equation}

\end{lemma}
\begin{IEEEproof}
    The lower bound is from the Gilbert--Varshamov (GV) bound, where the size of the confusable set of $\bpi$ is bounded by $|\cB_t(\bpi)|\le {n \choose t} \cdot t!{n \choose t}$ for any $\bpi\in \cS_n$. Hence, we have
    \begin{equation*}
        M(n,t) \ge \frac{n!}{{n \choose t} \cdot t!{n \choose t}}.
    \end{equation*}
    
    For the upper bound, let $\cD_t(\cS_n)=\cup_{\bpi\in\cS_n} \cD(\bpi)$. We have $|\cD_t(\cS_n)|=n!/t!$ since $\cD_t(\cS_n)$ is the set of all sequences consisting of $n-t$ distinct symbols from $\cS_n$. Applying the sphere-packing method, we have
    \begin{align*}
        &\frac{n!}{t!}\ge M(n,t)\cdot {n \choose t}.
    \end{align*}
    This completes the proof.\qedhere
\end{IEEEproof}

\begin{corollary}
    The lower bound of the minimal redundancy of the permutation codes for correcting $t$ deletions is $t\log n$.
\end{corollary}

\begin{lemma}[Section IV-C, \cite{gabrys2015codes}]\label{lem:gabrys_codesize}
Let $\cC_n$ be a permutation code for correcting $t$ deletions with length $n$. The size of the code $\cC_n$ is 
\begin{equation*}
    |\cC_n|\ge \frac{(\frac{2n}{t}!)^{\frac{t}{2}}}{(n+1)^{\frac{3t}{2}-4}(\frac{2n}{t})^{\frac{t}{2}}}.
\end{equation*}
\end{lemma}

The redundancy of the permutation code $\cC_n$ for correcting $t$ deletions proposed in \cite{gabrys2015codes} is far from $t\log n$, which is the lower bound of the minimal redundancy of the permutation code for correcting $t$ deletions.

\section{Improved GV Bound}\label{sec:improved_gv}


In this section, we achieve a logarithmic improvement on the GV bound of $M(n,t)$. The improved GV bound is presented as the following theorem.

\begin{theorem}\label{thm-main}
 For $n\ge t\geq 1$ be fixed positive integers, then 
 $$M(n,t) \geq \Omega_t\left(\frac{n! \log n}{n^{2t}}\right).$$
\end{theorem}
The proof idea of Theorem \ref{thm-main} relies on techniques from graph theory by constructing an auxiliary graph on the set of permutations such that each $t$-deletion permutation code is an independent set of this graph. Therefore, studying the size of codes is equivalent to analyzing the independence number of the auxiliary graph. It is well-known that if a graph has few triangles, we can improve the independence number by a log factor, compared with the classical bound by the greedy algorithm. Mention that this idea was initially used by Jiang and Vardy \cite{jiang2004asymptotic} to improve GV bound for binary codes with Hamming distance, followed by subsequent papers in other settings, see \cite{alon2023logarithmically, gao2013improvement,tait2013asymptotic} for examples. However, counting triangles is more challenging in our graph than in the highly symmetric setting studied by Jiang and Vardy \cite{jiang2004asymptotic}, where the corresponding graph is simply a power of the Hamming cube. In our case, the graph $\cG_{n,t}$, defined below, is not even regular, as some sequences of transportations can yield the same permutation from a given permutation. These challenges also arise in the recent work of Alon et al. \cite{alon2023logarithmically}, where they study $q$-ary codes with the deletion metric. They restricted attention to “pseudorandom” words
in the graph that are related by “pseudorandom” sequences of insertion and deletion operations, for suitable
notions of pseudorandomness. However, their approach does not carry over to our setting with permutations, since each insertion step depends on the previous deletion. Otherwise, there would be $n$ possible options for this operation, whereas in their case, the number of options is limited to $q$, which can be treated as a constant. To overcome these difficulties, we focus on the symbols used in each transportation instead of their positions, as the choice of symbols depends on the specific permutation.

Particularly, we demonstrate the proof of Theorem \ref{thm-main} through a two-step process: First, we simplify the issue by focusing on the enumeration of triangles within the $t$-deletion graph, denoted as $\cG_{n,t}$. Following that, we approximate the count of these triangles.
To achieve it, we shall use the following standard lemma of Bollob\'{a}s, which states that graphs with few triangles have large independence numbers. This lemma is a generalization of a result of Ajtai, Koml\'{o}s, and Szemer\'{e}di \cite{ajtai1980note} on triangle-free graphs. 


\begin{lemma}[Lemma 15~\cite{bollobas1998random}, p.296]\label{lemma-graph}
Let $G$ be a graph with maximum degree $\Delta$ ($\Delta \geq 1$) and suppose that $G$ has $T$ triangles. Then 
\begin{equation}
    \alpha(G) \geq \frac{|V(G)|}{10\Delta}\left(\log \Delta - \frac{1}{2}\log\left(\frac{T}{|V(G)|}\right)\right).
\end{equation}
where $\alpha(G)$ is the independent number of $G$ and $|V(G)|$ is the number of  vertices of $G$.
\end{lemma}

We define the graph $\cG_{n, t}$ with vertex set $\cS_n$, and two permutations are connected by an edge if they have a common subsequence of length at least $n - t$. Therefore, a $t$-deletion permutation code is an independent set of $\cG_{n, t}$. Note that $|V(\cG_{n,t})| = n!$ and $\Delta = O_{t}(n^{2t})$, since from any given permutation $\bsigma$, a neighbor $\bpi$ can be obtained by choosing $t$ positions of $\bsigma$ to delete in at most $\binom{n}{t}$ ways and then $t$ positions with letters to insert in at most $t!\binom{n}{t}$. Thus, in order to prove Theorem \ref{thm-main}, by using Lemma \ref{lemma-graph}, it suffices to show that the number of triangles in $\cG_{n, t}$ is $O_t(n! n^{4t - \varepsilon})$ for some $\varepsilon > 0$. We will prove this holds for $\varepsilon = t$. More precisely, we have the following lemma.
\begin{lemma}\label{lem.number-triangles}
 The number of triples $(\bsigma, \bpi, \btau) \in (\cS_n)^3$ such that $d_U(\bsigma,  \bpi) \leq t, d_U(\bpi, \btau) \leq t$, and $d_U(\bsigma, \btau) \leq t$ is at most $O_t(n! n^{3t})$. 
\end{lemma}

For $\bsigma \in \cS_n$, we will count the number of permutations $\bpi$ and $\btau$ such that $(\bsigma, \bpi, \btau)$ forms a triangle in $\cG_{n, t}$. Note that $(\bsigma, \bpi, \btau)$ is uniquely determined by $\bsigma$ and sequences $S_1 = (\phi_1,\phi_2,\dotsc, \phi_h), S_2 = (\phi'_1,\phi'_2, \dotsc \phi'_\ell)$ of translocations for which $\bpi = \bsigma \phi_1\phi_2\dotsm\phi_h$ and $\btau = \bpi\phi'_1\phi'_2\dotsm\phi'_\ell$. For the sake of brevity, we denote $\bpi\circ S=\bpi\phi_1\phi_2\dotsm\phi_m$, where the translocation sequence is $S=(\phi_1,\phi_2,\dotsc,\phi_m)$.


Since $d_U(\bsigma, \bpi) \leq t$ and $d_U(\bpi, \btau) \leq t$, we can choose $S_1, S_2$ such that $|S_1|, |S_2| \leq t$. Hence, we can combine $S_1$ and $S_2$ to obtain a sequence $S$ of translocations with size $|S_1| + |S_2| \leq 2t$ such that $\btau = \bsigma \circ S$. Now, for such a sequence $S$, there are at most $t = O_t(1)$ possibilities for $S_1$ and $S_2$ that produce $S$. Suppose $S = \phi_1 \phi_2 \dots \phi_m = S_1 \circ S_2$, with $|S_1| \leq t$ and $|S_2| \leq t$. Note that all translocation keep the same order and $|S_1| \leq t$ hence $S_1 = \phi_1 \phi_2 \dots \phi_k$ for some $k \leq t$ and $S_2$ is uniquely determined by $S_1$, i.e $S_2 = \phi_{k+1} \phi_{k+2} \dots \phi_m$. There are at most $t$ possibilities for $k$ such that $k \leq t$ and $m - k \leq t$, hence there are at most $t$ possibilities for $S_1$ and $S_2$. When $m = t+1$, $t$ possibilities can be achieved. Therefore, for a given $\bsigma$, the number of triangles $(\bsigma, \bpi, \btau)$ is at most $O_t(1)$ times the number of ways to pick a translocation sequence $S$ with $|S| \leq 2t$ such that $\btau = \bsigma \circ S$ and $d_U(\bsigma, \btau) \leq t$. Hence, Lemma \ref{lem.number-triangles} is followed by the following lemma. 

\begin{lemma}\label{lem:triangle}
  Given a permutation $\bsigma$, the number of translocation sequences $S$ with $|S| \leq 2t$ such that $\btau = \bsigma \circ S$ and $d_U(\bsigma, \btau) \leq t$ is at most $O_t(n^{3t})$. 
\end{lemma}


\begin{IEEEproof} 
Without loss of generality, we can assume that $\bsigma = \mathrm{id}$ and $|S| = 2t$, here $\mathrm{id}$ stands for identity permutation, i.e, $\mathrm{id}= (1, 2, \dots, n)$. We denote by $\varphi(a, j)$ the action that moves symbol $a$ to position $j$. This is equivalent to a translocation when we apply for a permutation. For example, if $\bpi = (\pi_1, \dots, \pi_n)$ and $\phi(i, j)$, then we have $\bpi \circ \phi(i, j) = \bpi \circ \varphi(\pi_i, j)$. Thus, we can consider a translocation sequence as a sequence of $\varphi(a, j)$. Let
$S =  \varphi(a_1, i_1) \circ \varphi(a_2, i_2)  \circ \dots \circ \varphi(a_{2t}, i_{2t})$ be a translocation sequence such that $d_U(\mathrm{id}, \mathrm{id} \circ  S)  \leq t$. This is equivalent to $\btau =  \mathrm{id} \circ S$ having an increasing subsequence of length at least $n - t$. Let $A = (a_1, \dots, a_{2t})$ be the sequence of symbols that are involved with translocations of $S$. Let $\Tilde{A}$ be the underlying set of $A$, i.e., the set of all distinct elements in $A$. For example, if $S = (\varphi(3, 2), \varphi(2, 5),\varphi(3, 8))$ then $A = (3, 2, 3)$ and $\Tilde{A} =\{2, 3\}$. We fall into the following cases.


\textbf{Case 1.} If $|\Tilde{A}| \leq t$, then there are at most $\binom{n}{\leq t} |\Tilde{A}|^{2t} = O_t(1) n^t$ possibilities for $A$. Combining with there are $n^{2t}$ choices for $i_1, \dots, i_{2t}$, the number of $S$ in this case is bounded by $O_t(1) n^{3t}$.

\textbf{Case 2.} If $|\Tilde{A}| = t + \ell$ for some $0 < \ell \leq t$. Let $C$ be the longest increasing subsequence of $\tau = \mathrm{id} \circ S$. Since $|C| \geq n - t$, $C$ contains at least $\ell$ elements in $A$. For an index $1  \leq j \leq 2t$, we say $j$ is \textit{good} if $a_j \in C$ and $a_j \notin \{a_{j+1}, \dots, a_{2t}\}$. In other words, $j$ is good if $\varphi(a_j, i_j)$ is the last time we move symbol $a_j$ in $S$. Note that after this translocation, the position of $a_j$ might be changed by other translocations but only at most $2t - j \leq 2t$ positions compared with $i_j$. 

Considering an \emph{good} index $j$, the position $x_j = \btau^{-1}(a_j)$ of $a_j$ in $\btau$ satisfies $i_j - 2t \leq x_j \leq i_j + 2t$. Since $a_j \in C$, we have
\begin{itemize}
    \item there are at least $a_j - t - 1$ elements of $\{1, 2, \dots, a_j - 1\}$ in $C$. If there are at most $a_j - t - 2$ elements of $\{1, 2, \dots, a_j - 1\}$ in $C$, then $|C| \leq (n - a_j + 1) + (a_j - t - 2) = n - t - 1$, which contradicts $|C| \geq n - t$. Therefore, the position of $a_j$ in $\tau$ is at least $a_j - t$, \textit{i.e} $x_j \geq a_j - t$. It follows that $i_j + 2t \geq x_j \geq a_j - t$, or $i_j \geq a_i - 3t$. 
    \item  there are at least $n - t - a_j$ elements of $\{a_j+1, \dots, n\}$ in $C$. Hence $i_j - 2t \leq x_j \leq a_j + t$, so $a_j \geq i_j - 3t$.
\end{itemize}
In short, for an arbitrary good index $j$, we have $a_j - 3t \leq i_j \leq a_j + 3t$. Clearly, we have at least $\ell$ good indices since there are at least $\ell$ elements of $\Tilde{A}$ that are contained in $C$.  This shows that 
\begin{equation}
  |S| \leq \binom{n}{t + \ell} (t+\ell)^{2t} \binom{t+\ell}{ \geq \ell} (6t + 1)^{\ell} \cdot n^{2t - \ell} = O_t(1) n^{3t},
\end{equation}
in which $\binom{n}{t + \ell} (t+\ell)^{2t}$ is the number of choices for $A$ and $\binom{t+\ell}{ \geq \ell}$ is the number of choices for good indices, and $n^{2t-\ell}$ is the number of positions that we can move \textit{not good} elements to, note that there are at most $2t - \ell$ not good element. 

\end{IEEEproof}

\section{Relationship of distances between two permutations}\label{sec:relation_distance}

In this section, we first build a tight inequality relationship between the Ulam distance and Hamming distance via a novel mapping function. Then we show that it is possible to construct permutation codes in various metrics using the known permutation code in the Hamming metric.
We note that permutation codes in the Hamming metric have been well-studied while there is a lack of knowledge in permutation codes in other metrics, such as the Ulam metric, generalized Kendall-$\tau$ metric, and generalized Cayley metric. We also show that our newly constructed permutation codes have better size than previously known permutation codes.


Farnoud, Skacheck, and Milenkovic~\cite{farnoud2013error} derived inequalities to present the relationship between the Ulam distance and Hamming distance over two permutations $\bpi$ and $\bsigma$. More precisely, they showed that 
\begin{equation}
    \frac{1}{n}d_H(\bpi,\bsigma)\le d_U(\bpi,\bsigma)\le d_H(\bpi,\bsigma), \text{ for all } \bpi, \bsigma \in \cS_n.
\end{equation}

In this paper, we propose a novel mapping function such that we can obtain a tighter inequality relationship between the Ulam distance and Hamming distance over two permutations.

Given a permutation $\bpi \in \cS_n$, we append $n+1$ at the end of $\bpi$ to obtain $\bar{\bpi}=(\bpi,n+1) \in \cS_{n+1}$. Let $\cS'_{n+1}$ be the subset of $\cS_{n+1}$ that consists of all permutations with the last element fixed as $n+1$, i.e. 
\[
\cS'_{n+1} = \left\{ \bar{\bpi} \in \cS_{n+1} \mid \bar{\bpi}_{n+1} = n+1 \right\}.
\]
Clearly, $\cS'_{n+1}$ is group isomorphic to $\cS_n$ and $|\cS'_{n+1}| = n!$.

\begin{definition}\label{def:mapping}
For $n \geq 1$, we define the function $\bsf:\cS_{n}\rightarrow \cS_{n+1}$, 
     such that $\bsf(\bpi) = (f(\bpi)_1, \dots, f(\bpi)_{n+1})$ where
     \begin{equation}\label{eq:mapping}
        f(\bpi)_i=\left\{
        \begin{array}{ll}
    \bar{\pi}_{\bar{\pi}^{-1}_i+1}, &  i=1,2,\dotsc,n \\
    \bar{\pi}_{1}, &  i=n+1
    \end{array}
    \right.
     \end{equation} 
for all $\bpi \in \cS_n$ in which $\bar{\bpi} = (\bpi, n+1) \in\cS'_{n+1}$. For convenience, we will write $\bsf(\bar{\bpi})$ instead of $\bsf(\bpi)$ and consider $\bsf$ as a function from $\cS'_{n+1}$ to $\cS_{n+1}$.
\end{definition}

\begin{example}
    Suppose $\bpi=(1,3,4,2,5)\in\cS_5$,  we have $\bar{\bpi}=(1,3,4,2,5,6)\in\cS'_6$ and $\bar{\bpi}^{-1}=(1,4,2,3,5,6)$. Hence,  $\bsf(\bar{\bpi})=(3,5,4,2,6,1)$.
\end{example}

One way we can think about $\bsf$ is the following. Starting with $\bpi$, we get $\bpi^{-1} = (\pi^{-1}_1, \dots, \pi^{-1}_n)$, increase all positions of $\bpi^{-1}$ by one, and append $1$ at the end to obtain a permutation $\balpha_{\pi}$ in $\cS_{n+1}$. Then $\bsf(\bpi)$ is obtained by taking the composition of $\bar{\bpi}$ and $\balpha$, which is denoted as $\bar{\bpi} \circ \balpha_{\pi}$. The process can be formally described as follows
\begin{align*}
    \bpi \to \bpi^{-1} = (\pi_1^{-1}, \dots, \pi_n^{-1}) \to \balpha_{\pi} := (1 + \pi_1^{-1}, \dots, 1 + \pi_n^{-1}, 1) \to \bsf(\bar{\bpi}) = \bar{\bpi} \circ \balpha_{\pi}.
 \end{align*}

Another more descriptive way to view $\bsf$ is that for a given permutation $\bar{\bpi} \in \cS'_{n+1}$, $\bsf$ maps elements $i$ to the elements that follow it in the sequence form of $\bar{\bpi}$, here the following element of $\bar{\pi}_{n+1}$ is $\bar{\pi}_1$. For example, suppose $\bar{\bpi} = (2,1,3,4,5)$ then $\bsf(1) = 3$ as $1$ is followed by $3$,  $\bsf(3) = 4$ as $3$ is followed by $4$ in $\bar{\bpi}$, and $\bsf(5) = 2$ for similar reason. 
From this viewpoint, it is easy to see that $\bsf$ is injective. For the sake of completeness, we include a formal proof below. 

\begin{claim}\label{claim:injective}
The function $\bsf$ is injective. 
\end{claim}

\begin{IEEEproof}
Suppose there are two permutations \(\bar{\bpi}\) and \(\bar{\bsigma}\) in \(\cS'_{n+1}\) such that \(\bsf(\bar{\bpi}) = \bsf(\bar{\bsigma})\), i.e. 
$f(\bar{\bpi})_i = f(\bar{\bsigma})_i$ for all $i = 1, 2, \ldots, n+1.$ By the definition of $\bsf$, we have $\bar{\pi}_1 = f(\bar{\bpi})_{n+1} = f(\bar{\bsigma})_{n+1} = \bar{\sigma}_1.$ Now, if $\bar{\pi}_k = \bar{\sigma}_k$ for some $1 \leq k < n+1$, we have 
\[ \bar{\pi}_{k+1} = \bar{\pi}_{\bar{\pi}^{-1}_i+1} = f(\bar{\bpi})_i = f(\bar{\bsigma})_i = \bar{\sigma}_{\bar{\sigma}^{-1}_i+1} = \bar{\sigma}_{k+1} \]
where $i =\bar{\pi}_k = \bar{\sigma}_k$. Therefore, it follows by induction that $\bar{\pi}_k = \bar{\sigma}_k$ for all $k = 1, 2, \dots, n+1$. In other words, $\bar{\bpi}  =\bar{\bsigma}$. 
This proves Claim \ref{claim:injective}.
\end{IEEEproof}

Let $f(\cS'_{n+1})$ be the image of $\cS'_{n+1} $ under the function  $\bsf$, namely 
$$f(\cS'_{n+1}) := \{ \bsf(\bar{\bpi}) \,|\,\bar{\bpi} \in \cS'_{n+1}\}.$$
By Claim \ref{claim:injective}, we have $|f(\cS'_{n+1})|=|\cS'_{n+1}| = |\cS_n| = n!$ and the inverse function  $\bsf^{-1}(\cdot) \colon f(\cS'_{n+1})\rightarrow\cS'_{n+1}$ is well-defined. Moreover, for any given $\bsigma\in f(\cS'_{n+1})$, the permutation $\bsf^{-1}(\bsigma) = (f^{-1}(\bsigma)_1, \dots, f^{-1}(\bsigma)_{n+1})$ can be recursively computed from $\bsigma$. In particular, we have $f^{-1}(\bsigma)_1 = \sigma_{n+1}$ and 
\begin{equation*}
    f^{-1}(\bsigma)_i =  \sigma_{f^{-1}(\bsigma)_{i-1}} \text{ for }  i= 2,3,\dots,n+1. 
\end{equation*}

Our main result of this section is the following theorem, showing a relationship between the Ulam distance over $\bpi$ and $\bsigma$ and the Hamming distance over two permutations $\bsf(\bpi)$ and $\bsf(\bsigma)$ for any $\bpi,\bsigma\in \cS_n$.

\begin{theorem}\label{thm:ulamhamming}
    Given two permutations $\bpi,\bsigma\in \cS_n$, we have $d_{U}(\bpi,\bsigma)\ge \frac{1}{3}d_{H}(\bsf(\bar{\bpi}),\bsf(\bar{\bsigma}))$.
\end{theorem}

\begin{IEEEproof}
Firstly, we consider the case when $d_U(\bpi, \bsigma)  = 1$, which means there is a translocation $\phi(i,j)$ such that $\bsigma = \bpi \phi(i,j)$. Without loss of generality, we may assume $1 \leq i < j \leq n$ since we can consider $\bpi = \bsigma \phi(j,i)$ if otherwise. Suppose $\bpi = (\pi_1, \dots, \pi_n)$, we have 
\begin{equation}\label{eq::sigma-pi}
  \bsigma = \bpi \phi(i,j) = (\pi_1, \dots, \pi_{i-1}, \pi_{i+1}, \dots, \pi_{j}, \pi_i, \pi_{j+1}, \dots, \pi_n).  
\end{equation} 
We recall that for a given permutation $\bar{\bpi}$, $f(\bar{\bpi})_i$ is the element that follow $i$ in $\bar{\bpi}$. Hence
\[ f(\bar{\bpi})_{\bar{\pi}_{n+1}} = \bar{\pi}_1 \text{ and } f(\bar{\bpi})_{\bar{\pi}_i} = \bar{\pi}_{i+1} \text{ for all } i = 1, \dots, n.\]
Similarly, it follows from \eqref{eq::sigma-pi} that $f(\bar{\bsigma})_{\bar{\pi}_{n+1}} = \bar{\pi}_1, f(\bar{\bsigma})_{\bar{\pi}_k} = \bar{\pi}_{k+1}$ if $k \notin \{i-1, i,  j\}$, and 
\begin{align*}
f(\bar{\bsigma})_{\bar{\pi}_{i-1}}=\pi_{i+1}, f(\bar{\bsigma})_{\bar{\pi}_{i}}=\pi_{j+1}, f(\bar{\bsigma})_{\bar{\pi}_{j}}=\pi_{i}.
\end{align*}
This shows that $\bsf(\bar{\bpi})$ and $\bsf(\bar{\bsigma})$ only differ at three positions $\bar{\pi}_{i-1}, \bar{\pi}_{i}$, and $\bar{\pi}_{j}$. Hence $d_H(\bsf(\bar{\bpi}), \bsf(\bar{\bsigma})) = 3$.  

Now, we suppose that $d_U(\bpi, \bsigma) = k$. By definition, there are $k$ translocations $\phi_i$ such that $\bsigma = \bpi \phi_1 \phi_2 \dots \phi_k$. Let $\bsigma^i = \bpi \phi_1 \dots \phi_i$ for $i = 1, \dots, k$. We have $\bsigma^{i} = \bsigma^{i-1} \phi_i$ for all $i = 1, \dots, k$ where $\bsigma^0 = \bpi$ and $\bsigma^k = \bsigma$. By above argument, we have $d_H(\bsf(\bar{\bsigma}^i), \bsf(\bar{\bsigma}^{i-1})) = 3$ for $i = 1, \dots, k$. It is followed by triangle inequalities that 
\[ d_H(\bsf(\bar{\bpi}), \bsf(\bar{\bsigma})) \leq d_H(\bsf(\bar{\bsigma}^0), \bsf(\bar{\bsigma}^1)) + d_H(\bsf(\bar{\bsigma}^1), \bsf(\bar{\bsigma}^2)) + \dots + d_H(\bsf((\bar{\bsigma}^{k-1}),\bsf(\bar{\bsigma}^k)) = 3k = 3 d_U(\bpi, \bsigma). \]
This completes the proof of Theorem \ref{thm:ulamhamming}.
\end{IEEEproof}

\begin{example}
    Suppose $\bpi=(1,3,4,2,5)$, $\bsigma=(4,2,3,5,1)$ and $d_U(\bpi,\bsigma)=2$, we have $\bar{\bpi}=(1,3,4,2,5,6)$ and $\bar{\bsigma}=(4,2,3,5,1,6)$. Then, we have $\bsf(\bar{\bpi})=(3,5,4,2,6,1)$ and $\bsf(\bar{\bsigma})=(6,3,5,2,1,4)$. Hence $d_H(\bsf(\bar{\bpi}),\bsf(\bar{\bsigma}))=5\le 3d_U(\bpi,\bsigma)$.
\end{example}

In the rest of this section, we shall construct permutation codes for correcting $t$ deletions by using Theorem~\ref{thm:ulamhamming} and some known permutation codes in the Hamming metric.
\begin{lemma}[Theorem 7, \cite{gabrys2015codes}]\label{lem:code_ryan}
    The permutation code $\cC\subseteq\cS_n$ is capable of correcting $t$ deletions if and only if $d_{U}(\bpi,\bsigma)> t$, for all $\bpi,\bsigma\in\cC$, $\bpi\neq\bsigma$.
\end{lemma}

\begin{theorem}\label{thm:codeinequal}
    Let $\cC\subseteq\cS_n$ be a permutation code where  $d_{H}(\bsf(\bar{\bpi}),\bsf(\bar{\bsigma}))\ge 3t+1$, for all $\bpi,\bsigma\in\cC$, $\bpi\neq\bsigma$. Then, $\cC$ is capable of correcting $t$ deletions.
\end{theorem}
\begin{IEEEproof}
This is immediately followed by combining Theorem~\ref{thm:ulamhamming} and Lemma \ref{lem:code_ryan}.
\end{IEEEproof} 

The construction of permutation codes in the Hamming metric is well-studied in \cite{gao2013improvement,jin2015construction,wang2017new}. Here, we apply the permutation code proposed in the Hamming metric in \cite{jin2015construction} as the base code. Let $A(n+1, d)$ be the maximum size of a permutation code of length $n+1$ with minimum Hamming distance $d$. Jin \cite{jin2015construction} showed that 
\begin{equation}\label{eq:per-H}
A(n+1,d) \geq \frac{(n+1)!}{p^{d-2}}    
\end{equation} 
where $p$ is the smallest prime bigger than or equal to $n$. Actually, the author obtained the inequality \eqref{eq:per-H} by proving the following stronger result.
\begin{lemma}[Theorem 2, \cite{jin2015construction}]\label{lem:lingfei_code}
For $n+1 \geq d \geq 4$, let $p$ be the smallest prime bigger than or equal to $n+1$. There exists a family $\{\cP_{i}(n+1, d_H): i=1,\ldots,p^{d_H-2} \}$ of $p^{d_H-2}$ permutation codes of length $n+1$ with minimum Hamming distance $d_H$ such that 
\[ \bigcup_{i=1}^{p^{d_H-2}} \cP_{i}(n+1,d_H) =\cS_{n+1}.\]
\end{lemma}

The following theorem presents the size of a permutation code with length $n$ that is capable of correcting $t$ deletions. 

\begin{theorem}
\label{cor:code_size}
    There exists a permutation code with length $n$ that is capable of correcting $t$ deletions with the code size at least $\frac{n!}{p^{3t-1}}$, where $p$ is the smallest prime bigger than or equal to $n$. Hence, it follows that the redundancy of this permutation code is at most $(3t-1)\log n+o(\log n)$ bits.
\end{theorem}

\begin{IEEEproof}
Let $d_H = 3t+1$ and let $\{\cP_{i}(n+1, d_H): i=1,\ldots,p^{d_H-2} \}$ be the family of $p^{d_H-2}$ permutation codes in Lemma~\ref{lem:lingfei_code}. For $i=1,2,\ldots,p^{d_H-2},$ let $\cP'_{i}(n+1,d_H)=\cP_{i}(n+1,d_H) \cap f(\cS'_{n+1})$ and 
    let $\cC_i(n, d_H)$ be the subset of $\cS_n$ that is obtained from $\bsf^{-1}\big( \cP'_{i}(n+1,d_H) \big)$ by removing the last symbol $n+1$ for each permutation. Since $\bigcup_{i=1}^{p^{d_H-2}} \cP_{i}(n+1,d_H) =\cS_{n+1}$ and $\bsf(\cS'_{n+1}) \subset \cS_{n+1}$, we obtain $\bigcup_{i=1}^{p^{d_H-2}} \bsf^{-1}(\cP'_{i}(n+1,d_H) ) =\cS'_{n+1}$ and thus $\bigcup_{i=1}^{p^{d_H-2}} \cC_i(n,d_H) =\cS_n.$ By pigeonhole principle, there exists $i \in \{1, \dots, p^{d_H-2}\}$ such that $\cC_i(n, d_H)$ is of size at least $\frac{n!}{p^{d_H-2}} = \frac{n!}{p^{3t-1}}$. On the other hand, it follows from Theorem \ref{thm:codeinequal} that $\cC_i(n, d_H)$ is a permutation code that is capable of correcting $t$ deletions. This completes the proof of Theorem~\ref{cor:code_size}.
\end{IEEEproof}

Furthermore, we can build the following relationship between the Levenshtein distance and Hamming distance over two permutations.
\begin{corollary}
 Given two permutations $\bpi,\bsigma\in \cS_n$, we have $ d_{H}(\bsf(\bar{\bpi}),\bsf(\bar{\bsigma}))\le \frac{3}{2}d_{L}(\bpi,\bsigma)$.
\end{corollary}

In addition, we consider establishing the relationship between the generalized Kendall-$\tau$ distance and the Hamming distance, as well as the relationship between the generalized Cayley distance and the Hamming distance. This is a natural consequence of Theorem~\ref{thm:ulamhamming}. Based on the definition, it follows that
\begin{equation*}
    d_{\bar{K}}(\bpi,\bsigma)\le d_{U}(\bpi,\bsigma)\le d_{K}(\bpi,\bsigma).
\end{equation*}

\begin{corollary}
 Given two permutations $\bpi,\bsigma\in \cS_n$, we have
 \begin{align*}
     d_{H}(\bsf(\bar{\bpi}),\bsf(\bar{\bsigma}))&\le 3d_{\bar{K}}(\bpi,\bsigma),\\
     d_{H}(\bsf(\bar{\bpi}),\bsf(\bar{\bsigma}))&\le 4d_{C}(\bpi,\bsigma),    
 \end{align*}
\end{corollary}
\begin{IEEEproof}
    The proof is almost identical to the proof of Theorem~\ref{thm:ulamhamming}. For the generalized Kendall-$\tau$ distance, $\bsf(\bar{\bpi})$ and $\bsf(\bar{\bsigma})$ differs at three positions $\bar{\pi}_{i-1}, \bar{\pi}_{j}$, and $\bar{\pi}_{\ell}$ if there is a single generalized adjacent transposition $\tau_{a}([i,j],[j+1,\ell])$. For the generalized Cayley distance, $\bsf(\bar{\bpi})$ and $\bsf(\bar{\bsigma})$ differs at four positions $\bar{\pi}_{i-1}, \bar{\pi}_{j}$, $\bar{\pi}_{k-1}$ and $\bar{\pi}_{\ell}$ if there is a single generalized adjacent transposition $\tau_{g}([i,j],[k,\ell])$. 
\end{IEEEproof}

Consequently, by leveraging Theorem~\ref{thm:ulamhamming} and its corollaries, we simplify the intricate task of developing permutation codes in the Levenshtein, Ulam, generalized Kendall-$\tau$, and generalized Cayley metrics. This simplification is achieved by transforming the problem into the construction of permutation codes in the Hamming metric, a well-studied area.

It is worth noting that in \cite{yang2019theoretical}, the authors constructed permutation codes in the generalized Cayley metric by establishing the connection $d_{B}(\bpi,\bsigma) \le 4d_{C}(\bpi,\bsigma)$ for any $\bpi,\bsigma \in \cS_n$, where $d_{B}(\bpi,\bsigma)$ is the block permutation distance proposed in \cite{yang2019theoretical}. However, the alphabet of the block permutation code is the smallest prime larger than $n^2 - n$, and the code size is at least $\frac{n!}{(2n^2-2n)^{4t-1}}$. Consequently, the size of the code proposed in \cite{yang2019theoretical} is smaller than that of our codes, which is at least $\frac{n!}{(2n)^{4t-1}}$. It is obtained by directly choosing the permutation code in the Hamming metric as the base code, which is similar to Theorem~\ref{cor:code_size} and its proof.

\section{Permutation Codes for Correcting $t$ deletions}\label{sec:construction}
In this section, we aim to design permutation codes for correcting $t$ deletions with a decoding algorithm. 
While the best-known permutation codes correcting $t$ deletions \cite{gabrys2015codes} are based on the auxiliary codes in the Ulam metric, our construction is achieved by incorporating the base code in the Hamming metric. Let $\cC_H(n,d_H)$ be a permutation code with minimum Hamming distance $d_H$. 
It is known 
that $\cC_H(n,d_H)$ can correct $t_1$ substitutions and $t_2$ erasures if $d_H > 2t_1+t_2$~\cite{gabrys2015codes}.  Let $\cD_H(n,t_1,t_2)$ be a decoder of the code $\cC_H(n,d_H)$ which can correct $t_1$ substitutions and $t_2$ erasures. That is, if the original codeword is a permutation of length $n$, $\bpi \in \cS_n$, and the input of the decoder $\cD_H(n,t_1,t_2)$ is a sequence obtained from $\bpi \in \cS_n$ after at most $t_1$ deletions and $t_2$ erasures, the output of the decoder $\cD_H(n,t_1,t_2)$ is the original permutation $\bpi \in \cS_n$. 

We now propose a permutation code correcting $t$ deletions with a decoding algorithm as follows.

\begin{theorem}\label{thmwithdecoding}
    For two integers $t,n$ with $n+1\geq 3t+1\ge 4$, let $\cC_H(n+1, 3t+1)$ be a permutation code of length $n+1$ with minimum Hamming distance $3t+1$. 
Then, the permutation code
    \begin{equation*}
        \cP_t(n)=\left\{\bpi\in \cS_n: \bsf(\bar{\bpi})\in\cC_{H}(n+1, 3t+1)\cap \bsf(\cS'_{n+1})\right\}
    \end{equation*}
    is capable of correcting $t$ deletions in $\bpi$.
\end{theorem}

\begin{IEEEproof}
    Although it is possible to obtain the result in Theorem \ref{thmwithdecoding} using the claim in Theorem \ref{thm:codeinequal}, we now focus on presenting a decoding algorithm in this proof. Let $\bpi =(\pi_1,\pi_2,\ldots,\pi_n) \in \cP_t(n)$ be the original permutation and let $\bsigma=(\sigma_1,\sigma_2,\ldots,\sigma_{n-t})$ be a sequence obtained from $\bpi$ after $t$ deletions. We also denote $\bar{\bsigma}=(\bsigma,n+1)$. Let $\bsf'(\bar{\bsigma})=(f_1,f_2,\ldots,f_n,f_{n+1})$ be a sequence of length $n+1$ such that for each $i \in [n+1]$:
    \begin{equation*}
    f_i = 
    \begin{cases} 
    \sigma_{j+1} & \text{if } i\in[n] \text{ and } \sigma_j = i, \text{ for some } j \in [n-t],\\
    * & \text{if } i\in[n] \text{ and } \sigma_j \neq i, \text{ for all } j \in [n-t],\\
    \sigma_1 & \text{if } i = n+1.
    \end{cases}
    \end{equation*}
    Similar to the proof of Theorem \ref{thm:ulamhamming}, the sequence $\bsf'(\bar{\bsigma})$ can be obtained from $\bsf(\bar{\bpi})$ after at most $t$ substitutions and $t$ erasures. Specifically, we first denote the set $I_t=\{d_1,\dotsc,d_t\}$ containing the indices of all $t$ deleted symbols. We notice that $f'(\bar{\bsigma})_{\pi_{d_1}},f'(\bar{\bsigma})_{\pi_{d_2}},\dotsc,f'(\bar{\bsigma})_{\pi_{d_t}}$ are marked as $*$ in $\bsf'(\bar{\bsigma})$, which can be considered as $t$ erasures when comparing $\bsf'(\bar{\bsigma})$ and $\bsf(\bar{\bpi})$. Then, we have $\bsf'(\bar{\bsigma})_{\pi_{d_1-1}}\neq \bsf(\bar{\bpi})_{\pi_{d_1-1}},\dotsc, \bsf'(\bar{\bsigma})_{\pi_{d_t-1}}\neq \bsf(\bar{\bpi})_{\pi_{d_t-1}}$ if $\{d_1-1,\dotsc,d_t-1\}\cap I_t=\emptyset$. It also means that there are at most $t$ substitutions between $\bsf'(\bar{\bsigma})$ and $\bsf(\bar{\bpi})$.  Since $\bpi \in \cP_t(n)$ and $\bsf(\bar{\bpi})$ can correct $t$ substitutions and $t$ erasures, the decoder $\cD_H(n+1,t,t)$ can recover the sequence $\bsf(\bar{\bpi})$ from the sequence $\bsf'(\bar{\bsigma})$. Furthermore, since the mapping $\bsf$ is injective, it is possible to obtain the original permutation $\bpi$ from the sequence $\bsf(\bar{\bpi})$. Hence, the permutation code $\cP_t(n)$ constructed above can correct $t$ deletions. 
\end{IEEEproof}
    
\begin{example}
    Let $\bpi=(1,3,4,2,5,6,9,8,7)\in\cS_9$ and $t=2$ with $(3,8)$ are deleted. Then, we have the following
    \begin{align*}
        \bar{\bsigma}=(1,4,2,5,6,9,7,10),\quad \bsf'(\bar{\bsigma})=(\underline{4},5,*,2,6,9,10,*,\underline{7},1),\\
        \bar{\bpi}=(1,3,4,2,5,6,9,8,7,10),  \quad \bsf(\bar{\bpi})=(3,5,4,2,6,9,10,7,8,1).
    \end{align*}
    where symbols with underlined denote the symbols are substituted and $*$ denotes the erasure. We can see that there are $2$ substitutions and $2$ erasures between $\bsf'(\bar{\bsigma})$ and $\bsf(\bar{\bpi})$ when $t=2$.
\end{example}

From the above proof of Theorem \ref{thmwithdecoding}, we obtain a decoding algorithm of $\cP_t(n)$. The details of the decoding algorithm will be described in Algorithm \ref{alg:decoding}.

\begin{algorithm}
\SetAlgoLined
\KwInput{Sequence $\bsigma$ obtained from $\bpi$ after $t$ deletions}
\KwOutput{Original permutation $\bpi\in\cP_t(n)$}
Comparing $\bsigma$ and $[n]$, find the set $A_t$ of $t$ deleted symbols.

Define the sequence $\bsf'(\bar{\bsigma})=(f_1,f_2,\ldots,f_n,f_{n+1})$ such that for each $i \in A_t$ then $f_i = *$ and for the rest of $i \in [n]$, find $\sigma_j=i$ and assign $f_i=\sigma_{j+1}$. Also, let $f_{n+1}=\sigma_1$.
  
Use the decoder $\cD_H(n+1,t,t)$ with the input $\bsf'(\bar{\bsigma})$ to obtain the sequence $\bsf(\bar{\bpi})$.

Recover the original permutation $\bpi$ by finding the reverse mapping $\bsf^{-1}$ of $\bsf(\bar{\bpi})$ and remove the last $n+1$.

 \caption{Decoding Algorithm of $\cP_t(n)$}\label{alg:decoding}
\end{algorithm}

We note that the size of the code $\cP_t(n)$ is dependent on the permutation code $\cC_H(n+1,3t+1)$. As we have shown in Theorem~\ref{cor:code_size}, if we choose a permutation code from \cite{jin2015construction}, we can obtain a permutation code correcting $t$ deletions with size at least $\frac{n!}{p^{3t-1}}$, where $p$ is the smallest prime larger than $n$.
The code size of our proposed permutation codes for correcting $t$ deletions greatly improves the results in \cite{gabrys2015codes} as shown in Lemma~\ref{lem:gabrys_codesize}. 
Besides that, the decoding algorithm of $\cP_t(n)$ in Algorithm~\ref{alg:decoding} is efficient. In particular, the decoding complexity is dominated by Step 3 in Algorithm 1 which is dependent on the decoder $\cD_H(n+1,t,t)$ of the permutation code $\cC_H(n+1,3t+1)$. If the decoder of a permutation code in the Hamming metric is linear complexity then the decoder of our constructed permutation code is also linear complexity.

\section{Permutation codes for correcting multiple bursts of deletions}\label{sec:multiburst}
In this section, we consider the problem of constructing codes that can correct at most $t$ bursts of deletions each of length at most $b$ occur. Formally, a \emph{burst of at most $b$ deletions} deletes at most $b$ consecutive symbols from the permutation $\bpi$, leading to $\bpi'=(\pi_1,\pi_2,\dotsc,\pi_i,\pi_{i+b'+1},\dotsc,\pi_n)$, where $b'\leq b$. Although the binary codes~\cite{lenz2020optimal}, non-binary codes~\cite{wang2023non}, and permutation codes~\cite{Wang2022permutation,sun2023improved} for correcting a burst of $b$ deletions can achieve the order-optimal redundancy, only a few works study the code for correcting multiple bursts of deletions. Sima et al.~\cite{sima2020syndrome} propose binary codes for correcting at most $t$ bursts of deletions with each of length at most $b$ with $4t(1+\epsilon)\log n$ bits of redundancy. Recently, Ye et al.~\cite{ye2024codes} studied the binary codes for correcting two bursts of \emph{exactly} $b$ deletions with $5\log n+O(\log\log n)$ bits of redundancy. To the best of our knowledge, this is the first trial to construct non-binary/permutation codes for correcting at most $t$ bursts of deletions with each of length at most $b$.

\begin{theorem}\label{multiburst}
    For integers $t,b,n$ with $n+1\geq 2t+tb+1\ge 4$, let $\cC_H(n+1, 2t+tb+1)$ be a permutation code of length $n+1$ with minimum Hamming distance $2t+tb+1$. 
Then, the permutation code
    \begin{equation*}
        \cP_t^b(n)=\left\{\bpi\in \cS_n: \bsf(\bar{\bpi})\in\cC_{H}(n+1, 2t+tb+1)\cap \bsf(\cS'_{n+1})\right\}
    \end{equation*}
    is capable of correcting at most $t$ bursts of deletions with each of length at most $b$ in $\bpi$.
\end{theorem}

\begin{IEEEproof}
    The main idea of the proof is similar to that of Theorem~\ref{thmwithdecoding}. Let $\bpi =(\pi_1,\pi_2,\ldots,\pi_n) \in \cP_t^b(n)$ be the original permutation and let $\bsigma=(\sigma_1,\sigma_2,\ldots,\sigma_{n-tb})$ be a sequence obtained from $\bpi$ after $t$ bursts of deletions with each of length $b$. We also denote $\bar{\bsigma}=(\bsigma,n+1)$. Let $\bsf'(\bar{\bsigma})=(f_1,f_2,\ldots,f_n,f_{n+1})$ be a sequence of length $n+1$ such that for each $i \in [n+1]$:
    \begin{equation*}
    f_i = 
    \begin{cases} 
    \sigma_{j+1} & \text{if } i\in[n] \text{ and } \sigma_j = i, \text{ for some } j \in [n-tb],\\
    * & \text{if } i\in[n] \text{ and } \sigma_j \neq i, \text{ for all } j \in [n-tb],\\
    \sigma_1 & \text{if } i = n+1.
    \end{cases}
    \end{equation*}
    Specifically, we first denote the set $I_t=\{d_1^1,\dotsc,d_1^b,d_2^1,\dotsc,d_2^b,\dotsc,d_t^1,\dotsc,d_t^b\}$ containing the indices of all $tb$ deleted symbols. We notice that $f'(\bar{\bsigma})_{\pi_{d_1^1}},\dotsc,f'(\bar{\bsigma})_{\pi_{d_1^b}},\dotsc,f'(\bar{\bsigma})_{\pi_{d_t^1}},\dotsc,f'(\bar{\bsigma})_{\pi_{d_t^b}}$ are marked as $*$ in $\bsf'(\bar{\bsigma})$, which can be considered as $tb$ erasures when comparing $\bsf'(\bar{\bsigma})$ and $\bsf(\bar{\bpi})$. Then, we have $\bsf'(\bar{\bsigma})_{\pi_{d_1^1-1}}\neq \bsf(\bar{\bpi})_{\pi_{d_1^1-1}},\dotsc, \bsf'(\bar{\bsigma})_{\pi_{d_t^1-1}}\neq \bsf(\bar{\bpi})_{\pi_{d_t^1-1}}$ if $\{d_1^1-1,\dotsc,d_t^1-1\}\cap I_t=\emptyset$. It also means that there are at most $t$ substitutions between $\bsf'(\bar{\bsigma})$ and $\bsf(\bar{\bpi})$.  Since $\bpi \in \cP_t(n)$ and $\bsf(\bar{\bpi})$ can correct $t$ substitutions and $tb$ erasures, the decoder $\cD_H(n+1,t,tb)$ can recover the sequence $\bsf(\bar{\bpi})$ from the sequence $\bsf'(\bar{\bsigma})$. Furthermore, since the mapping $\bsf$ is injective, it is possible to obtain the original permutation $\bpi$ from the sequence $\bsf(\bar{\bpi})$. Hence, the permutation code $\cP_t^b(n)$ constructed above can correct at most $t$ burst of deletions with each of length at most $b$. 
\end{IEEEproof}
\begin{example}
    Let $\bpi=(1,3,4,2,5,6,9,8,7)\in\cS_9$ and $t=2, b=2$ with $(3,4)$ and $(9,8)$ are deleted. Then, we have the following
    \begin{align*}
        \bar{\bsigma}=(1,2,5,6,7,10),\quad &\bsf'(\bar{\bsigma})=(\underline{2},5,*,*,6,\underline{7},10,*,*,1),\\
        \bar{\bpi}=(1,3,4,2,5,6,9,8,7,10),  \quad &\bsf(\bar{\bpi})=(3,5,4,2,6,9,10,7,8,1).
    \end{align*}
    where symbols with underlined denote the symbols are substituted and $*$ denotes the erasure. We can see that there are $t=2$ substitutions and $tb=4$ erasures between $\bsf'(\bar{\bsigma})$ and $\bsf(\bar{\bpi})$ when $t=2,b=2$.
\end{example}

For the size of the code $\cP_t^b(n)$, as we have shown in Section~\ref{sec:construction}, if we choose a permutation code from \cite{jin2015construction}, we can obtain a permutation code correcting $t$ deletions with size at least $\frac{n!}{p^{2t+tb-1}}$, where $p$ is the smallest prime larger than $n$. Hence, the redundancy of the permutation code for correcting at most $t$ bursts of deletions with each of length at most $b$ is at most $(2t+tb-1)\log n+o(\log n)$ bits.

\section{Multipermutation codes for correcting $t$ deletions}\label{sec:multipermutation}

In this section, we will extend the permutation codes for correcting $t$ deletions to multipermutations. A \textit{multipermutation} is an ordered arrangement of elements from a multiset. Given a positive integer \( n \) and a multiplicity vector \( \br = (r_1, r_2, \dots, r_k) \), such that \( n = \sum_{i=1}^k r_i \), we denote the multiset as \( \cM^{\br}_n \). 

\[
\cM^{\br}_n = \{ \underbrace{1, \ldots, 1}_{r_1}, \underbrace{2, \ldots, 2}_{r_2}, \ldots, \underbrace{k, \ldots, k}_{r_k} \}.
\]

A multiset where each element appears exactly \( r \) times is called an \textit{\( r \)-regular multiset}. For brevity, we denote the \( r \)-regular multiset as \( \cM^r_n \). An \textit{\( r \)-regular multipermutation} is defined as any permutation of an \( r \)-regular multiset. Throughout this paper, we focus on \( r \)-regular multipermutations.

Let $n = r \cdot k$, where $r \geq 1$ and $k \geq 1$. Consider the set $\mathcal{M}_n^r$ of $r$-regular multipermutations of length $n$ over the symbol set $ \{1,2,\dotsc,k\}$, where each symbol appears exactly  $r$ times.
We extend $\bx\in\cM_n^r$ by appending $r$ copies of $k+1$ at the end, resulting in: 
\begin{equation*}
    \bar{\bx} = (x_1, x_2, \dotsc, x_n, \underbrace{k+1, k+1, \dotsc, k+1}_{\text{\( r \) times}})\in \cM_{n+r}^r.
\end{equation*}
Let $\bar{\cM}_{n+r}^r$ be the subset of $\cM_{n+r}^r$ that consists of all multipermutations with the last $r$ elements fixed as $k+1$, i.e.
\[
 \bar{\cM}_{n+r}^r= \left\{ \bar{\bx} \in \cM_{n+r}^r \mid \bar{\bx}_{[n+1:n+r]} = k+1 \right\}.
\]
Clearly, $\bar{\cM}_{n+r}^r$ is group isomorphic to $\cM_{n}^r$ and $|\bar{\cM}_{n+r}^r|=\frac{n!}{(r!)^{n/r}}$.

\begin{definition}
    For $n,k,r\ge 1$, we define the mapping \( \mathbf{g}: \mathcal{M}_n^r \rightarrow \mathcal{M}_{n+r}^r \) such that $\bg(\bx) = (g(\bx)_1, \dots, g(\bx)_{n+r})$ where:

\begin{equation}
    g(\bx)_s = \begin{cases}
\bar{x}_{p_{a,\ell}+1}, & \text{if } p_{a,\ell}+1 \leq n+r, \\
\bar{x}_{1}, & \text{if } p_{a,\ell}+1 > n+r,
\end{cases}
\end{equation}
for \( s = (\ell - 1) \cdot (k+1) + a \), where  $a \in \{1,2,\dotsc,k+1\}$, $\ell \in \{1,2,\dotsc,r\}$, and $p_{a,\ell}$ denotes the position of the $\ell$-th occurrence of symbol $a$ in $\bx$. For convenience, we will write $\bg(\bar{\bx})$ instead of $\bg(\bx)$ and consider $\bg$ as a function from $\bar{\cM}_{n+r}^r$ to $\cM_{n+r}^r$.
\end{definition}
\begin{example}
    Suppose $\bx=(4,1,3,1,2,2,3,4)\in\cM^2_{8}$, we have $\bar{\bx}=(4,1,3,1,2,2,3,4,5,5)$. Also, we have $p_{1,1}=2, p_{2,1}=5, p_{3,1}=3, p_{4,1}=1, p_{5,1}=9,  p_{1,2}=4, p_{2,2}=6, p_{3,2}=7, p_{4,2}=8, p_{5,2}=10$. Hence, $\bg(\bar{\bx})=(3,2,1,1,5,2,3,4,5,4)$.
\end{example}

\begin{lemma}\label{lem:multi_relation}
        Given two multipermutations $\bx,\by\in \cM_n^r$, we have $d_{U}(\bx,\by)\ge \frac{1}{r+2}d_{H}(\bg(\bar{\bx}),\bg(\bar{\by}))$.
    \end{lemma}

    \begin{IEEEproof}
    We firstly consider the case when $d_U(\bx, \by)  = 1$, which means there is a translocation $\phi(i,j)$ such that $\by = \bx \phi(i,j)$. Without loss of generality, we may assume $1 \leq i < j \leq n$ since we can consider $\bx = \by \phi(j,i)$ if otherwise. Suppose $\bx = (x_1, \dots, x_n)\in\cM_n^r$, we have 
    \begin{equation}\label{eq::sigma-pi}
        \by = \bx \phi(i,j) = (x_1, \dots, x_{i-1}, x_{i+1}, \dots, x_{j}, x_i, x_{j+1}, \dots, x_n).  
    \end{equation} 
Similar to the proof in Theorem~\ref{thm:ulamhamming}, we consider the differences between  $\bg(\bar{\bx})$ and $\bg(\bar{\by})$ due to the translocation:
    \begin{enumerate}
        \item Mapping for \( x_i \): The indexes of $x_i$ in $\bar{\bx}$ and $\bar{\by}$ are given by $i=p_{x_i,m}$ and $j=p_{x_i,m'}$, where $m$ and $m'$ denote the $m$-th and $m'$-th occurrence of symbol $x_i$ in $\bar{\bx}$ and $\bar{\by}$, respectively. Here, $m\leq m'$ since $i\leq j$. In the worst case, we can observe when $\bar{m}\in[m,m'-1]$, $\bg(\bar{\bx})_{s_{\bar{m}}}=\bar{x}_{p_{x_i,\bar{m}}+1}$ but $\bg(\bar{\by})_{s_{\bar{m}}}=\bar{x}_{p_{x_i,\bar{m}+1}+1}$, where $s_{\bar{m}}=(\bar{m}-1)(k+1)+x_i$. When $\bar{m}=m'$, we have $\bg(\bar{\bx})_{s_{\bar{m}}}=\bar{x}_{p_{x_i,m'}+1}$ but $\bg(\bar{\by})_{s_{\bar{m}}}=\bar{x}_{j+1}$. Hence, we can conclude that
        \begin{equation*}
            \bg(\bar{\bx})_s\neq \bg(\bar{\by})_s, \text{for}\; s=(\bar{m}-1)(k+1)+x_i, \forall \bar{m}\in[m,m']
        \end{equation*}
        \item Mapping for \( x_{i-1} \): We observe that the translocation of $x_i$ does not influence the order of occurrence of $x_{i-1}$. We denote $i-1=p_{x_{i-1},m_1}$ and $s_{i-1}=(m_1-1)(k+1)+x_{i-1}$. Then, we have $\bg(\bar{\bx})_{s_{i-1}}\neq \bg(\bar{\by})_{s_{i-1}}$ since $\bg(\bar{\bx})_{s_{i-1}}=\bar{x}_i$ but $\bg(\bar{\by})_{s_{i-1}}=\bar{x}_{i+1}$. If $i=1$, we consider the last $k+1$ as $x_{i-1}$.
       
        \item Mapping for \( x_j \): When $x_j=x_i$, $x_j$ is actually the $(m'-1)$-th occurrence of $x_i$ in $\bar{\by}$. As we claimed in Case 1, we already take into account all of the differences between   $\bg(\bar{\bx})_s$ and $\bg(\bar{\by})_s$, for $s=(\bar{m}-1)(k+1)+x_i, \forall \bar{m}\in[m,m']$. Next, when $x_j\neq x_i$, as in Case 2, the translocation of $x_i$ does not influence the order of occurrence of $x_{j}$. Hence, we have $\bg(\bar{\bx})_{s_{j}}\neq \bg(\bar{\by})_{s_{j}}$ since $\bg(\bar{\bx})_{s_{j}}=\bar{x}_{j+1}$ but $\bg(\bar{\by})_{s_{j}}=\bar{x}_{i}$, where $j=p_{x_{j},m_2}$ and $s_{j}=(m_2-1)k+x_{j}$
    \end{enumerate}

Therefore, we can see $\bg(\bar{\bx})$ and $\bg(\bar{\by})$ differ at most $r+2$ positions. Hence, $d_H(\bg(\bar{\bx}),\bg(\bar{\by}))=r+2$. Now, the same as the proof of Theorem~\ref{thm:ulamhamming}, we have $d_H(\bg(\bar{\bx}),\bg(\bar{\by}))\leq (r+2) d_U(\bx,\by)$. This completes the proof of this lemma.
\end{IEEEproof}
\begin{example}
    Suppose $\bx=(4,1,3,1,2,2,3,4)\in\cM^2_{8}$ and $\by=\bx\phi(2,6)=(4,3,1,2,2,1,3,4)$. Then, we have $\bg(\bar{\bx})=(3,2,1,1,5,2,3,4,5,4)$ and $\bg(\bar{\by})=(\underline{2},2,1,\underline{3},5,\underline{3},\underline{1},4,5,4)$. We can see $d_H(\bg(\bar{\bx}),\bg(\bar{\by})=4=r+2$. Specifically, the 2\textsuperscript{nd} and 4\textsuperscript{th} substitutions result from different mappings for $x_{i-1}$ and $x_j$, respectively. Meanwhile, the 1\textsuperscript{st} and 3\textsuperscript{rd} substitutions indicate that the translocation of $x_i$ not only affects the mapping of $x_i$ itself but also influences the mapping of any symbol identical to $x_i$ located within the index interval $[i, j]$.
   
\end{example}

Based on the relationship between Ulam distance and Levenshtein distance as shown in \eqref{eq:ulamleven}, we have 
\begin{equation*}
    d_H(\bg(\bar{\bx}),\bg(\bar{\by}))\leq (r+2) d_U(\bx,\by)=\frac{r+2}{2} d_L(\bx,\by)
\end{equation*}
We denote $\cC_t^r(n)\subseteq\cM_n^r$ as the \emph{$r$-regular multipermutation codes} for correcting $t$ deletions and  $\cC_d^r(n)\subseteq\cM_n^r$ as the \emph{$r$-regular multipermutation codes} with minimal Hamming distance $d$. Similar to Theorem~\ref{thm:codeinequal}, we have the following:
\begin{theorem}
    Let $\cC_t^r(n)\subseteq\cM_n^r$ be a permutation code where  $d_{H}(\bg(\bar{\bx}),\bg(\bar{\by}))\ge (r+2)t+1$, for all $\bx,\by\in\cC_t^r(n)$, $\bx\neq\by$. Then, $\cC_t^r(n)$ is capable of correcting $t$ deletions.
\end{theorem}

Hence, the $r$-regular multipermutation code $\cC_t^r(n)$ for correcting $t$ deletions can be achieved by applying the $r$-regular multipermutation codes with minimal Hamming distance $(r+2)t+1$ as the base code.
As for the specific construction for multipermutation codes $\cC_d^r(n)$ with minimal Hamming distance $d=(r+2)t+1$, we refer to \cite{luo2003constant,ding2005combinatorial,huczynska2006frequency}. For the code size, it was proved in \cite{hassanzadeh2014multipermutation} that:
\begin{equation}\label{eq:multihamming}
    |\cC_d^r(n)|\geq \frac{n!}{(r!)^{n/r}{n\choose d-1}(\frac{n}{r})^{d-1}}.
\end{equation}
Hence, 
\begin{equation*}
    |\cC_t^r(n)|\geq \frac{n!}{(r!)^{n/r}{n\choose (r+2)t}(\frac{n}{r})^{(r+2)t}}.
\end{equation*}
Furthermore, we define the redundancy of the $r$-regular multipermutation code $\cC_t^r(n)$ as $\log (|M_n^r|/|\cC_t^r(n)|)$. The redundancy of our $r$-regular multipermutation code for correcting $t$ deletions is at most $2(r+2)t\log n$ bits of redundancy when $r,t$ are constants. To the best of our knowledge, this is the first work to consider multipermutation codes for correcting multiple \textit{stable} deletions.  Sala et al.~\cite{sala2014deletions} proposed a multipermutation code for correcting $t$ \textit{unstable} deletions, where the decoder only examines ranks and cannot tell the absolute value of the deleted element. In the construction of this code, each symbol $x_i$ in the codeword should satisfy $x_i\equiv a_i \bmod (t+1)(2t+1)+1$. Hence, the redundancy of this multipermutation code for correcting $t$ unstable deletions is extremely high.   

It is also interesting to notice that an $r$-regular multipermutation code with minimal Ulam distance $d_u$ is shown in \cite{hassanzadeh2014multipermutation}, which is achieved by interleaving $r$ permutation code with minimal Ulam distance $d_u$. We denote this multipermutation code as $\cC_{\text{FM}}(n,r,d_u)$ with code size:
\begin{equation*}
    |\cC_{\text{FM}}(n,r,d_u)|\geq \left(\frac{(n/r-d_u+1)!}{{n/r\choose d_u-1}}\right)^r
\end{equation*}
From Lemma~\ref{lem:multi_relation} and \eqref{eq:multihamming}, we can obtain a new $r$-regular multipermutation code with minimal Ulam distance $d_u$, denoted as $\cC(n,r,d_u)$. The code size of $\cC(n,r,d_u)$ is:
\begin{equation*}
    |\cC(n,r,d_u)|\geq \frac{n!}{(r!)^{n/r}{n\choose (r+2)d_u}(\frac{n}{r})^{(r+2)d_u}}.
\end{equation*}
\begin{claim}\label{claim:compare}
For $r\geq 2, d_u\ge 2$ and $\delta> 0$, we have $\frac{|\cC_{\text{FM}}(n,r,d_u)|}{ |\cC(n,r,d_u)|}=e^{-n\delta}$ as  $n\to \infty$, when $r$ and $d_u$ are constants.
\end{claim}
The proof of this claim is provided in Appendix~\ref{app:compare}. Therefore, this new $r$-regular multipermutation code $\mathcal{C}(n, r, d_u)$, constructed by utilizing the multipermutation code in the Hamming metric as the base code, achieves a much larger code size.

\section{Conclusion}\label{sec:conclusion}
Motivated by various applications of permutation codes, we studied the theoretical bound and construction of permutation codes in Levenshtein, Ulam, and Generalized Kendall-$\tau$ metrics in this paper. We achieve a logarithmic improvement on the GV bound of the maximum size of the $t$-deletion correcting permutation codes. We then provide a construction of permutation codes correcting $t$ deletions with an efficient decoding algorithm. We also show permutation codes for correcting multiple bursts of deletions and multipermutation codes for correcting $t$ deletions. Our constructed codes are better than the previously known results. However, the constructed code is not a systematic code. Hence, in future work, we will aim to design a systematic code with an efficient encoding/decoding algorithm. 
Besides that, there are many avenues for future research including constructing permutation codes for correcting $2$ deletions with a lower redundancy. 

\section*{Acknowledgement}
The second author would like to thank Jozsef Balogh, Quy Dang Ngo, and Ethan White for useful discussions. The Nguyen was partially supported by a David G. Bourgin Mathematics Fellowship.

\bibliographystyle{IEEEtran}
\bibliography{references}

\begin{appendices}
\section {Proof of Claim~\ref{claim:compare}}\label{app:compare}
\begin{IEEEproof}
    We first denote $|\cC_{\text{FM}}(n,r,d_u)|$ and $|\cC(n,r,d_u)|$ as $E_1$ and $E_2$ respectively for simplification.
    We have:
    \[
    \log E_1 = r \left[ \log  \left( \dfrac{ n }{ r } - d_u + 1 \right)!  - \log \dbinom{ n/r }{ d_u - 1 } \right].
    \]
Also, 
    \[
\log \dbinom{ n/r }{ d_u - 1 } = \log \left( \dfrac{ n }{ r } ! \right) - \log ( d_u - 1 )! - \log \left( \dfrac{ n }{ r } - d_u + 1 \right)!.
\]
Then, 
\[
\begin{aligned}
\log E_1 &= r \left[ \log \left(  \dfrac{ n }{ r } - d_u + 1 \right)! - \left( \log \left( \dfrac{ n }{ r } ! \right) - \log ( d_u - 1 )! - \log \left( \dfrac{ n }{ r } - d_u + 1 \right)! \right) \right] \\
&= r \left[ 2 \log  \left( \dfrac{ n }{ r } - d_u + 1 \right)!  - \log \left( \dfrac{ n }{ r } ! \right) + \log ( d_u - 1 )! \right].
\end{aligned}
\]
We apply Stirling's approximation and let \( N = \dfrac{ n }{ r } \), so \( N \to \infty \) as \( n \to \infty \). Then, we have:
\[
\begin{aligned}
\log \left( N ! \right) &\approx N \log N - N, \\
\log \left( N - d_u + 1 \right)! &\approx \left( N - d_u + 1 \right) \log \left( N - d_u + 1 \right) - \left( N - d_u + 1 \right).
\end{aligned}
\]
Hence,
\[
\begin{aligned}
\log E_1 &\approx r \left[ 2 \left( N - d_u + 1 \right) \log \left( N - d_u + 1 \right) - 2 \left( N - d_u + 1 \right) - N \log N + N + \log ( d_u - 1 )! \right].
\end{aligned}
\]
Since \( d_u \) is a constant and \( N \to \infty \), we can approximate as $\log \left( N - d_u + 1 \right) \approx \log N$.
We have
\[
\begin{aligned}
\log E_1 \approx r \left[ N \log N - N -2(d_u-1)\log N\right]+ \text{constant}.
\end{aligned}
\]
We recall that \( r N = n \) and \( \log N = \log n - \log r \), then 
\begin{equation}
\log E_1 \approx n \log n - n ( 1 + \log r )- 2r(d_u-1)\log n + \text{constant}.
\end{equation}

On the other side, we have
\[
\log E_2 = \log n! - \frac{ n }{ r } \log ( r! ) - \log \dbinom{ n }{ ( r + 2 ) d_u } - ( r + 2 ) d_u \log \left( \dfrac{ n }{ r } \right).
\]
Term $\log n!$:
\[
\log n! \approx n \log n - n.
\]
Term $\frac{ n }{ r } \log ( r! )$:
\[
\frac{ n }{ r } \log ( r! ) = n C_1, \quad \text{where } C_1 = \frac{ \log ( r! ) }{ r }.
\]
Term $\log \dbinom{ n }{ ( r + 2 ) d_u }$:
\[
\log \dbinom{ n }{ ( r + 2 ) d_u } = \log n! - \log \left( ( r + 2 ) d_u \right)! - \log \left( n - ( r + 2 ) d_u \right)!.
\]
Applying Stirling's approximation:
\[
\begin{aligned}
\log n! &\approx n \log n - n, \\
\log \left( ( r + 2 ) d_u \right)! &\approx ( r + 2 ) d_u \log \left( ( r + 2 ) d_u \right) - ( r + 2 ) d_u, \quad (\text{constant}) \\
\log \left( n - ( r + 2 ) d_u \right)! &\approx \left( n - ( r + 2 ) d_u \right) \log \left( n - ( r + 2 ) d_u \right) - \left( n - ( r + 2 ) d_u \right).
\end{aligned}
\]
Since \( ( r + 2 ) d_u \) is constant and \( n \to \infty \), we have $
n - ( r + 2 ) d_u \approx n$ and  $\log \left( n - ( r + 2 ) d_u \right) \approx \log n.
$
Then,
\[
\begin{aligned}
\log \left( n - ( r + 2 ) d_u \right)! &\approx \left( n - ( r + 2 ) d_u \right) \log n - \left( n - ( r + 2 ) d_u \right).
\end{aligned}
\]
By combining terms:
\[
\log \dbinom{ n }{ ( r + 2 ) d_u } \approx  ( r + 2 ) d_u \log n + \text{constant}.
\]
Term $( r + 2 ) d_u \log \left( \dfrac{ n }{ r } \right )$:
\[
( r + 2 ) d_u \log \left( \dfrac{ n }{ r } \right ) = ( r + 2 ) d_u ( \log n - \log r ) = ( r + 2 ) d_u \log n + \text{constant}.
\]
Then, sum up all the above terms:
\begin{equation*}
\log E_2 \approx  n \log n - n ( 1 + C_1 ) - 2( r + 2 ) d_u \log n + \text{constant}.
\end{equation*}

Next, we have
\begin{equation*}
    \log E_1-\log E_2=(C_1-\log r)n+2( r + 2 d_u) \log n.
\end{equation*}
The dominant term here is $(C_1-\log r)n$. Then, when $r\geq 2$, we recall that \( C_1 = \dfrac{ \log ( r! ) }{ r } \) and denote:
\[
\delta = \log r - \frac{ \log ( r! ) }{ r } > 0.
\]
Therefore, for \( \delta > 0 \), we have
\[
\log E_1 - \log E_2 \approx - n \delta.
\]
This implies:
\[
\frac{ E_1 }{ E_2 } = e^{ - n \delta } \to 0 \quad \text{as } n \to \infty.
\]
This completes our proof.
\end{IEEEproof}
\end{appendices}

\end{document}